\documentclass[unsortedaddress,preprint,show pacs]{revtex4-1}
\usepackage{graphicx}
\usepackage{dcolumn}
\usepackage{amsmath}
\usepackage{amssymb}
\usepackage{amsfonts}
\usepackage{bm}
\usepackage{color}
\newcommand{\be}{\begin{equation}}
\newcommand{\ee}{\end{equation}}
\newcommand{\ba}{\begin{eqnarray}}
\newcommand{\ea}{\end{eqnarray}}

\begin{document}
\title{Probing the existence of phase transitions\\in one-dimensional fluids of penetrable particles}
\author{Santi Prestipino$^{1,2}$\footnote{Corresponding author. Email: {\tt sprestipino@unime.it}}, Domenico Gazzillo$^3$\footnote{Email: {\tt gazzillo@unive.it}}, and Nicola Tasinato$^3$\footnote{Email: {\tt tasinato@unive.it}}}
\affiliation{$^1$Universit\`a degli Studi di Messina, Dipartimento di Fisica e di Scienze della Terra, Contrada Papardo, I-98166 Messina, Italy\\$^2$CNR-IPCF, Viale F. Stagno d'Alcontres 37, I-98158 Messina, Italy\\$^3$Universit\`a Ca' Foscari Venezia, Dipartimento di Scienze Molecolari e Nanosistemi, S. Marta DD 2137, I-30123 Venezia, Italy}
\date{\today }
\begin{abstract}
Phase transitions in one-dimensional classical fluids are usually ruled out by making appeal to van Hove's theorem. A way to circumvent the conclusions of the theorem is to consider an interparticle potential that is everywhere bounded. Such is the case of, {\it e.g.}, the generalized exponential model of index 4 (GEM-4 potential), which in three dimensions gives a reasonable description of the effective repulsion between flexible dendrimers in a solution. An extensive Monte Carlo simulation of the one-dimensional GEM-4 model [S. Prestipino, {\em Phys. Rev. E} {\bf 90}, 042306 (2014)] has recently provided evidence of an infinite sequence of low-temperature cluster phases, however also suggesting that upon pushing the simulation forward what seemed a true transition may eventually prove to be only a sharp crossover. We hereby investigate this problem theoretically, by three different and increasingly sophisticated approaches ({\it i.e.}, a mean-field theory, the transfer matrix of a lattice model of clusters, and the exact treatment of a system of point clusters in the continuum), to conclude that the alleged transitions of the one-dimensional GEM4 system are likely just crossovers.
\end{abstract}
%   61.20.Gy: Theory and models of liquid structure
%   64.60.A-: Specific approaches applied to studies of phase transitions
%   64.60.De: Statistical mechanics of model systems
\maketitle

\section{Introduction}
Compared to three dimensions, the statistical mechanics of one-dimensional (1D) fluids is both simpler and harder. Simpler, because a general result due to van Hove~\cite{vanHove} excludes the possibility of distinct phases in one-dimensional, homogeneous, and one-component classical fluids of hard particles with short-range interactions. Also more complicated, however, because whenever any of the hypotheses behind van Hove's theorem is not met, it may be extremely difficult to say whether or not the given 1D interaction would admit a genuine phase transition (see, \textit{e.g.}, Refs.\thinspace \cite{Cuesta,Cuesta2,Ares,Fantoni,Santos,Speranza1,Giaquinta,Pekalski}).

In this respect, the class of potentials known as generalized-exponential models (GEM) is emblematic. These are classical fluids of particles repelling each other through a potential of the form $u(r)=\epsilon\exp\{-(r/\sigma )^{\alpha }\}$ (GEM-$\alpha $ potential), where $\epsilon,\sigma >0$ are arbitrary energy and length units, $r$ is the interparticle distance, and $\alpha$ is a real positive index. The GEM potentials stay finite at the origin, hence particles may fully overlap, in strike contrast to real atoms and molecules. Clearly, bounded potentials only make sense as models of the \emph{effective} pair interaction between large molecular aggregates that can intertwine even to the extent that their centers of mass exactly coincide. In three dimensions bounded interactions give rise to structural patterns which are unknown to rare-gas fluids, notably reentrant melting and waterlike anomalies, which have been the subject of numerous investigations in the last few decades (see, \textit{e.g.}, Refs.~\cite{Likos1,Mausbach,Krekelberg,Prestipino5,Speranza2}, to name but a few). A particularly curious feature of bounded two-body repulsions is the possibility to generate stable \emph{cluster crystals}\thinspace\cite{Likos2}, \textit{i.e.}, crystalline phases involving a mean number of particles per lattice site larger than 1, which have been extensively studied in the recent past\thinspace\cite{Likos3,Mladek1,Mladek2,Zhang1,Zhang2,Zhang3,Neuhaus,Wilding1,Prestipino4}.

When confined in one dimension, penetrable particles may pass one through the other, thus making it possible to change the particle ordering within the line. While there exists a general method of computing the exact thermodynamics of a specific class of 1D systems of impenetrable particles\thinspace \cite{Takahashi,Gursey,Lieb}, nothing can \emph{a priori} be said about the phase behavior of freely overlapping particles. Hence, either a brute-force numerical-simulation approach is taken or some other plausibility arguments must be used to probe the existence of separate phases in such 1D systems.

In Ref.\thinspace\cite{Speranza1}, the 1D GEM-2 (that is, the one-dimensional counterpart of the Gaussian-core model\thinspace \cite{Stillinger,Lang,Prestipino1,Prestipino2}) was studied by Monte Carlo (MC) simulation. No phase transition was found, even at zero temperature ($T=0$), notwithstanding the system structure at low temperature is crystal-like and clear signs of dimerization were found at high pressure. For the 1D GEM-4 the situation is markedly different\thinspace\cite{Prestipino3}. True phase transitions exist at $T=0$ between states allowing for an increasing number of particles in the same lattice site. In simulation, these states gave rise to clear-cut cluster phases for $T>0$, with sharp boundaries which manifest themselves through hysteresis loops in the number density as a function of pressure. In fact, the coexistence loci are so sharp that they could be traced by free-energy calculations. However, when MC simulations were made much longer (up to tens of million sweeps), hysteresis loops became narrower and the alleged critical points shifted to low temperatures, thus raising doubts on the true nature of the crossover from one phase to the other. Therefore, it is still unclear whether proper phase transitions truly exist or not in the 1D GEM-4.

In the present paper, we address this problem from the different perspective of theoretical modeling. Upon building up successive, increasingly accurate approximations to the 1D GEM-$\alpha $ system for $\alpha\ge 4$, we compute the density as a function of pressure for fixed temperature, seeking for jumps or other singularities. At the end, we get a more definite idea about the nature of phase crossovers in GEM fluids. Even though not having the same degree of certainty as a mathematical proof, nonetheless the present work sheds fresh light on a still open question.

This paper is organized as follows. After introducing a coarse-grained formulation of the 1D GEM-$\alpha $ Hamiltonian in terms of interacting clusters on a lattice, we compute the phase boundaries of the model first in the mean-field approximation (Section II), and then by the exact transfer-matrix method (Section III). Going back to the continuum, in Section IV we analyze the exact behavior of a system of point clusters moving on a line and discuss its relevance for the phase behavior of the original GEM system. Some concluding remarks are given in the last section.

\section{Mean-field theory}
\setcounter{equation}{0}
\renewcommand{\theequation}{2.\arabic{equation}}

The statistical mechanics of the 1D GEM-$\alpha $ system is not amenable to an exact analytical treatment. However, some progress can be made by analyzing a related system which allows to simplify the counting of states. In this regard, a crucial observation from the simulation~\cite{Prestipino3} is that the low-temperature structure of the GEM fluid away from the alleged \textquotedblleft transition points\textquotedblright\ is not dissimilar from an ordered cluster crystal, with average site occupancy and lattice spacing depending on temperature $T$ and number density $\rho $ in a smooth way. At a phase transition, the average occupancy increases by one and the density undergoes a sharp variation (however, close to a transition point clusters do not necessarily contain the same number of particles and the separation between two adjacent clusters may then vary along the line). Given the above evidence, we are led to represent the low-temperature behavior of the 1D GEM-$\alpha $ system of $N$ particles in a volume $L$ by means of the coarse-grained lattice model of Hamiltonian: 
\begin{equation}
H=\frac{1}{2}\epsilon \sum_{i=1}^{N_{\mathrm{c}}}n_{i}(n_{i}-1)+\sum_{i=1}^{N_{\mathrm{c}}}n_{i}\sum_{m>0}u\left( m\,r_{\mathrm{NN}}\right)n_{i+m}\,,
\label{2-1}
\end{equation}
where $N_{\mathrm{c}}$ is the (for the moment arbitrary) number of point-like clusters (one for each lattice site), $r_{\mathrm{NN}}=L/N_{\mathrm{c}}$ is the lattice spacing, \textit{i.e.}, the distance between nearest-neighbor (NN) clusters, and $n_{i}=1,2,3,\ldots $ is the size of the cluster residing at $i$, \textit{i.e.}, the number of particles located at the fixed lattice positions 
\begin{equation}
x_{i}=(i-1)r_{\mathrm{NN}}\,\,\,\,\,\,(i=1,\ldots ,N_{\mathrm{c}})\,.
\label{2-2}
\end{equation}
On the right-hand side of Eq.\thinspace (\ref{2-1}), the first term is the intra-cluster energy while the second term represents the total interaction energy of particles belonging to different clusters (periodic boundary conditions are implied). Observe that the inter-cluster potential in Eq.\thinspace (\ref{2-1}) is not restricted to nearest-neighbors, but involves arbitrarily distant clusters. A model similar to (\ref{2-1}) has already been considered by Wilding and Sollich~\cite{Wilding2} as a tool to investigate the $T=0$ transitions of the three-dimensional GEM-$\alpha$ system. In this section we extend their mean-field treatment to $T>0$ for the 1D case, while deferring the exact analysis of model (\ref{2-1}) to the next section.

In order to extract the thermodynamic properties of (\ref{2-1}) we first use the variational principle of statistical mechanics. Since handling the $\sum_in_i=N$ constraint is hard, we shall work in the grand-canonical ensemble where $T,L$, and the chemical potential $\mu$ are fixed, while each $n_i$ can assume every value between 1 and $\infty$. In this ensemble the Gibbs-Bogoliubov inequality prescribes that the grand potential $\Omega$ is bounded from above by a functional $\Omega^*$,
\begin{equation}
\Omega\le\Omega^*[\pi]\equiv\mathrm{Tr}\left[\pi(H+\frac{1}{\beta}\ln\pi-\mu N)\right]\equiv E^*-TS^*-\mu N^*\,,
\label{2-3}
\end{equation}
where $\pi$ is any trial probability density in state space and $\beta=(k_{\mathrm{B}}T)^{-1}$. $\Omega^*$ reaches its minimum for $\pi=\rho_{\mathrm{g.\,c.}}$ (\textit{i.e.}, the equilibrium grand-canonical density), where it takes the value $\Omega$. Once a parametric form of $\pi$ is chosen, minimizing $\Omega^*[\pi]$ with respect to parameters is the way to make the best possible estimate of $\Omega$.

A system with no specific constraint on the total number of lattice sites entails a complication. In fact, after computing $\overline{\Omega}$ ({\it i.e.}, the least upper bound to $\Omega$) for each $N_{\mathrm{c}}$, we still need to optimize with respect to $N_{\mathrm{c}}$. This is done according to the criterion stated by Swope and Andersen~\cite{Swope}, that is by minimizing $\overline{\Omega }$ with respect to $r_{\mathrm{NN}}=L/N_{\mathrm{c}}$, or, equivalently, with respect to the $N_{\mathrm{c}}$-dependent density $d\equiv N^{\ast}/L$~\cite{note}.

The simplest choice is a $\pi (n_{1},n_{2},\ldots ,n_{N_{\mathrm{c}}})$ of uncorrelated $n_{i}$ variables. Following Ref.\thinspace \cite{Wilding2}, we choose $\overline{n}$ (\textit{i.e.}, the mean site occupancy) as the only parameter in $\pi $. At low temperature, we are allowed to assume that only two $n_{i}$ values at a time may occur in the system, namely $n_{<}=\mathrm{int}(\overline{n})$ and $n_{>}=n_{<}+1$. Hence we have: 
\begin{equation}
\pi(\{n\};\overline{n})=\prod_{i=1}^{N_{\mathrm{c}}}\pi _{1}(n_{i};\overline{n})\,,\,\,\,\,\,\,\mathrm{with}\,\,\pi _{1}(n;\overline{n})=(1-\Delta )\delta _{n,n_{<}}+\Delta \delta _{n,n_{>}}\,.
\label{2-4}
\end{equation}
The value of $\Delta $ is determined by the condition $\langle n_{i}\rangle_{\pi }\equiv \mathrm{Tr}(\pi n_{i})=\overline{n}$. Since
\begin{equation}
\langle n_{1}\rangle _{\pi }=\sum_{n_{1}}n_{1}\pi _{1}(n_{1};\overline{n})\times \sum_{n_{2}}\pi _{1}(n_{2};\overline{n})\times \cdots =(1-\Delta)n_{<}+\Delta n_{>}\,,
\label{2-5}
\end{equation}
we get $\Delta =\overline{n}-n_{<}$, which also yields: 
\begin{equation}
\langle n_{1}^{2}\rangle _{\pi }=(1-\Delta )n_{<}^{2}+\Delta n_{>}^{2}=\overline{n}^{2}+\Delta (1-\Delta )\,.
\label{2-6}
\end{equation}
From the very definition of $N^{\ast }$ it follows that $N^{\ast }=N_{\mathrm{c}}\overline{n}$, hence
\begin{equation}
N_{\mathrm{c}}=Ld/\overline{n}\,.
\label{2-7}
\end{equation}
Putting all things together we obtain: 
\begin{eqnarray}
E^{\ast } &=&\frac{Ld}{2\overline{n}}\left\{ \left[ \overline{n}^{2}+\Delta(1-\Delta )-\overline{n}\right] \epsilon +2\overline{n}^{2}\left[ u\left(\frac{\overline{n}}{d}\right) +u\left( \frac{2\overline{n}}{d}\right)+\ldots \right] \right\} \,;
\notag \\
-TS^{\ast } &=&k_{\mathrm{B}}T\frac{Ld}{\overline{n}}\left[ (1-\Delta )\ln(1-\Delta )+\Delta \ln \Delta \right] \,;
\notag \\
-\mu N^{\ast } &=&-\mu Ld\,,
\label{2-8}
\end{eqnarray}
and the system pressure $P$ is then given by: 
\begin{equation}
P(T,\mu )=-\frac{1}{L}\min_{d}\overline{\Omega }(d;T,\mu )=-\frac{1}{L}\min_{d}\left\{ \min_{\overline{n}}\Omega ^{\ast }(\overline{n},d;T,\mu)\right\} \,.
\label{2-9}
\end{equation}
Any jump in the equilibrium number density $\rho \equiv \mathrm{arg}\min_{d}\left\{ \overline{\Omega }(d;T,\mu )\right\} $ as a function of $\mu$ at constant $T$ will be the hallmark of a first-order transition.

In Figs.\thinspace 1 and 2 we have plotted our mean-field results for the 1D GEM-4 system relative to six isothermal paths, $T=0,0.01,\ldots ,0.05$ (from now on, all quantities will be given in the units set by $k_{\mathrm{B}},\epsilon $, and $\sigma $). For these temperatures we first computed the best value of $\overline{n}$ and the corresponding value $\overline{F}$ of $F^{\ast }=E^{\ast }-TS^{\ast }$ as a function of $d$ (top panel of Fig.\thinspace 1). The optimal $\overline{n}$ undergoes a steady increase with $d$, interrupted at regular intervals by plateaus at integer heights. Within the same $d$ ranges where $\overline{n}$ stays constant, $\overline{F}$ is convex upward. The bottom panel of Fig.\thinspace 1 shows the lattice spacing $r_{\mathrm{NN}}=L/N_{\mathrm{c}}=\overline{n}/d$ as a function of $d$; $r_{\mathrm{NN}}$ exhibits a zig-zag pattern which gradually dampens as $d$ grows. Observe that the value of $r_{\mathrm{NN}}$ never falls below $1.1$, meaning that the distance between two next-nearest-neighbor clusters is always larger than $2.2$. Since $\exp \{-2.2^{4}\}<10^{-10}$, this means that the GEM-4 interactions are effectively limited to nearest neighbors only (the same applies for any $\alpha >4$). In Fig.\thinspace 2 we show $\rho $ (top panel) and $\mu $ (bottom panel) as a function of $P$ for the same temperatures considered in Fig.\thinspace 1. The density jumps at the lowest temperatures indicate a sequence of first-order transitions between cluster phases of increasing occupancy. Above $T\approx 0.03$ all discontinuities vanish and the transitions are replaced by smooth crossovers.

Summing up, our mean-field theory asserts the existence of low-temperature phase transitions in the 1D GEM-4 system. Compared to the work by Wilding and Sollich\,\cite{Wilding2}, two important improvements are: 1) our theory is directly derived from the variational principle of statistical mechanics and, as such, it can be systematically improved; 2) at variance with the theory in Ref.\,\cite{Wilding2}, to which ours reduces for $T=0$, the present mean-field theory also applies for $T>0$.

\section{Transfer-matrix treatment}
\setcounter{equation}{0}
\renewcommand{\theequation}{3.\arabic{equation}}

Since the results of an approximate theory may not be conclusive, the accuracy of our mean-field theory for 1D GEM-$\alpha $ systems will now be tested against an exact treatment of model (\ref{2-1}). We start by rewriting the latter Hamiltonian as:
\begin{equation}
H=\frac{1}{2}\epsilon \sum_{i=1}^{N_{\mathrm{c}}}n_{i}(n_{i}-1)+\sum_{i=1}^{N_{\mathrm{c}}}u(r_{\mathrm{NN}})n_{i}n_{i+1}\,,
\label{3-1}
\end{equation}
which assumes zero interaction beyond nearest neighbors (we shall check \textit{a posteriori} that, for $\alpha \geq 4$, the omitted terms are actually ineffective at all temperatures of interest). Again, the clusters are fixed at the lattice positions given by Eq.\thinspace (\ref{2-2}), and $r_{\mathrm{NN}}=L/N_{\mathrm{c}}$ is an adjustable parameter to be eventually optimized by requiring the system pressure to be maximum. Each $N_{\mathrm{c}}$-uple of $n_{i}$ values (for $i=1,\ldots ,N_{\mathrm{c}}$) represents an individual state of the model (\ref{3-1}), since the particles building up the clusters are indistinguishable. The \emph{canonical} partition function would be the sum of the Boltzmann factors over all $N_{\mathrm{c}}$-uples $(n_{1},\ldots ,n_{N_{\mathrm{c}}})$ of positive integers obeying the constraint $\sum_{i}n_{i}=N$. Therefore, as in the previous section, the calculation is most easily performed in the grand-canonical ensemble, since the partition function of the latter involves an unconstrained sum over $n_{1},\ldots ,n_{N_{\mathrm{c}}}$.

Using the transfer-matrix method, the grand potential (under periodic boundary conditions) reads
\begin{eqnarray}
-\beta \Omega (T,L,\mu ;r_{\mathrm{NN}}) &=&\ln \mathrm{Tr}\left\{e^{\beta\mu N}e^{-\beta H}\right\} =\ln \sum_{n_{1},\ldots ,n_{N_{\mathrm{c}}}}{\mathbb{T}}_{n_{2}}^{n_{1}}{\mathbb{T}}_{n_{3}}^{n_{2}}\cdots {\mathbb{T}}_{n_{1}}^{n_{N_{\mathrm{c}}}}
\notag \\
&=&\ln \sum_{n_{1}}\left( {\mathbb{T}}^{N_{\mathrm{c}}}\right)_{n_{1}}^{n_{1}}\sim N_{\mathrm{c}}\ln \lambda _{\mathrm{max}}\,,
\label{3-2}
\end{eqnarray}
where the last step holds for large $N_{\mathrm{c}}$ (and $L$) values. In the above equation, $\lambda _{\mathrm{max}}$ is the largest eigenvalue of the (symmetric) transfer matrix 
\begin{equation}
{\mathbb{T}}_{n^{\prime }}^{n}=\exp \left\{ -\beta \epsilon \frac{n^{2}+n^{\prime 2}-n-n^{\prime }}{4}+\beta \mu (n+n^{\prime })/2-\beta u(r_{\mathrm{NN}})nn^{\prime }\right\} \,.
\label{3-3}
\end{equation}
It is worth stressing that the thermodynamic limit implies taking not only $L\rightarrow \infty $ but also $N_{\mathrm{c}}\rightarrow \infty $, in such a way that $L/N_{\mathrm{c}}=r_{\mathrm{NN}}$ remains finite. Also observe that ${\mathbb{T}}$ is an infinite matrix, since both $n$ and $n^{\prime }$ can vary between 1 and $\infty $. In practice, if the system density is not too high, we can assume $n$ and $n^{\prime }$ in Eq.\thinspace (\ref{3-3}) to vary between, say, 1 and 10 (${\mathbb{T}}$ thus becomes a $10\times 10$ matrix).

The calculation of $\lambda _{\mathrm{max}}$ was carried out numerically by the power method~\cite{Scott}. First the system pressure for every $r_{\mathrm{NN}}$ was computed through $\Omega (T,L,\mu ;r_{\mathrm{NN}})=-\mathcal{P}(T,\mu ;r_{\mathrm{NN}})L$, yielding in the thermodynamic limit:
\begin{equation}
\mathcal{P}(T,\mu ;r_{\mathrm{NN}})=\frac{\ln \lambda _{\mathrm{max}}}{\beta r_{\mathrm{NN}}}\,.
\label{3-4}
\end{equation}
Such a quantity was then maximized to get the equilibrium pressure: 
\begin{equation}
P(T,\mu )=\max_{r_{\mathrm{NN}}}\mathcal{P}(T,\mu ;r_{\mathrm{NN}})\,.
\label{3-5}
\end{equation}
A first-order phase transition occurs whenever the $\mathcal{P}(T,\mu ;r_{\mathrm{NN}})$ vs. $r_{\mathrm{NN}}$ profile exhibits two different maxima, whose relative stability changes when crossing a specific $\mu$ value. This implies a jump discontinuity in the optimal value of $r_{\mathrm{NN}}$, which in turn produces a jump in the density as a function of $\mu $ or $P$.

We first computed $P$ for $T=0.01,\ldots ,0.05$, finding values very close to those given by mean-field theory (pressure differences in reduced units were typically smaller than $10^{-3}$); only for the smallest $T$ the numerical calculation of the pressure could not be carried out above a certain $\mu $, because of the huge values of some ${\mathbb{T}}$ components. We verified that, even including second-neighbor terms in the Hamiltonian, the pressure does not change appreciably for $\alpha \geq 4$. The reason is the same mentioned previously: the optimal $r_{\mathrm{NN}}$ values are never smaller than $1.1\sigma $, hence the strength of the second-neighbor repulsions is nearly zero. In Fig.\thinspace 3 we show our exact transfer-matrix results for the lattice model (\ref{3-1}). Besides $r_{\mathrm{NN}}$, the top panel of Fig.\thinspace 3 also reports the connected correlation between NN clusters, $\langle n_{1}n_{2}\rangle _{\mathrm{c}}\equiv \langle n_{1}n_{2}\rangle -\langle n_{1}\rangle \langle n_{2}\rangle$, and the correlation length $\xi $ defined by 
\begin{equation}
\xi =\frac{1}{\ln \left\vert \lambda _{\mathrm{max}}/\lambda _{2}\right\vert}\,,
\label{3-6}
\end{equation}
where $\lambda _{2}$ is the second (next to the leading) eigenvalue of ${\mathbb{T}}$ in descending order. While $\lambda _{2}$ was computed by Wielandt deflation~\cite{Saad}, the NN correlation was determined by the formula: 
\begin{equation}
\langle n_{1}n_{2}\rangle =\frac{1}{\lambda _{\mathrm{max}}}\sum_{n_{1},n_{2}=1}^{10}{\mathbb{R}}_{1}^{n_{1}}{\mathbb{R}}_{1}^{n_{2}}{\mathbb{T}}_{n_{2}}^{n_{1}}n_{1}n_{2}\,,
\label{3-7}
\end{equation}
where the ${\mathbb{R}}$ matrix has the ${\mathbb{T}}$ eigenvectors for columns (in descending order of eigenvalue). The absolute value of $\langle n_{1}n_{2}\rangle_{\mathrm{c}}$, which is very small for all pressures, may explain the good quality of our mean-field theory. The bottom panel of Fig.\thinspace 3 reports the average occupancy $\langle n_{1}\rangle$, the average squared occupancy $\langle n_{1}^{2}\rangle$, and the density $\rho$, respectively computed by the formulas: 
\begin{equation}
\langle n_{1}\rangle =\sum_{n_{1}=1}^{10}\left( {\mathbb{R}}_{1}^{n_{1}}\right) ^{2}n_{1}\,,\,\,\,\langle n_{1}^{2}\rangle=\sum_{n_{1}=1}^{10}\left( {\mathbb{R}}_{1}^{n_{1}}\right)^{2}n_{1}^{2}\,,\,\,\,\mathrm{and}\,\,\,\rho =\langle n_{1}\rangle /r_{\mathrm{NN}}\,.
\label{3-8}
\end{equation}
Overall, the density behavior closely resembles that found in mean-field theory. Indeed, we see again discontinuous jumps at the lowest temperatures ($T=0,0.01,0.02$), indicating true first-order phase transitions. At higher temperatures, the discontinuous jumps transform into sharp, but continuous, density changes in a very narrow range of pressures (similar trends were named \emph{pseudo-transitions} in a different 1D model by P\c{e}kalski \emph{et al.}~\cite{Pekalski}).

The correlation length shows a distinct maximum near each low-temperature transition point. As $T$ grows from 0, these maxima first increase and then, past $T=0.03$, start to decrease, suggesting that each coexistence line ends with a critical point and that $\xi $ diverges at a critical temperature $\approx 0.03$. Since the transfer matrix ${\mathbb{T}}$ is strictly positive for any $T$, one might argue that its maximum eigenvalue is necessarily simple (by Perron-Frobenius theorem~\cite{Perron}), and therefore it would be safely excluded that $\xi $ may ever diverge. However, as already noted, the true transfer matrix is infinite and therefore the use of the Perron-Frobenius theorem is not legitimate~\cite{Evans}. Hence, the question of the behavior of $\xi $ near $T=0.03$ can only be answered numerically (see below).

Upon looking more deeply into the density and correlation-length profiles close to every zero-temperature transition point, we observed a very complex behavior. In Figs.\thinspace 4 and 5 we focus on the transition from the 1-cluster phase (1) to the 2-cluster phase (2). The density jump at the transition has an irregular $T$ behavior. Upon heating from $T=0$, the 1-2 coexistence line \textquotedblleft bifurcates\textquotedblright\ at $T\lesssim 0.028$. Indeed, two distinct density jumps are seen for $T=0.028$ (red curves in Figs.\thinspace 4 and 5), and this entails the existence of a third phase, intermediate between 1 and 2, characterized by values of $\overline{n}\lesssim 1.5$. However, this extra-phase has a very precarious existence, since it has already disappeared for $T=0.029$. Eventually, above $T=0.033$ we find no density jump but a continuous density curve, and the 1-2 coexistence line terminates. Figure 5 shows that the correlation length never diverges and that the Widom line (\textit{i.e.}, the locus of regular $\xi $ maxima), rather than departing from the ending point of the coexistence line, actually blossoms from the coexistence line near $T=0.028$. We finally notice that the behavior of $\rho $ and $\xi $ near the 2-3 and 3-4 transitions is perfectly analogous to that described for the 1-2 transition.

Figures 6 and 7 report the overall phase diagram on the $P$-$T$ and $\rho $-$T$ plane, respectively. On each coexistence line there is a bifurcation point and the Widom line apparently emerges from it. The mean-field coexistence lines are also shown. They run extremely close to the exact loci up to the bifurcation points, while progressively departing from them for higher temperatures.

Finally, we looked at what happens when increasing $\alpha$ from 4. We found that the discontinuous density jumps are only present in a smaller and smaller range of low temperatures until they altogether disappear for all $T>0$ when $\alpha\rightarrow\infty$. In this limit, $\langle n_1\rangle$ coincides with $\rho$ (see Fig.\,8). Hence, the 1D penetrable-sphere model (PSM, corresponding to the $\alpha\rightarrow\infty$ limit of GEM-$\alpha$) would show no phase transition at $T>0$.

\section{A continuum model of interacting clusters}
\setcounter{equation}{0}
\renewcommand{\theequation}{4.\arabic{equation}}

We hereafter show that in Hamiltonian (\ref{3-1}) the lattice constraint (\ref{2-2}) can be weakened, making the distance between clusters partly arbitrary, and the model still remains \emph{exactly} solvable.

In the previous section, we considered a model whose states are represented by $N_{\mathrm{c}}$-uples of positive $n_{i}$ integers, defined at fixed lattice positions $x_{i}$. Slightly abusing the notation, a model state was any vector $(n_{1},x_{1},\ldots ,n_{N_{\mathrm{c}}},x_{N_{\mathrm{c}}})$, with $x_{i}$ defined by Eq.\thinspace (\ref{2-2}). We now release the constraint on the cluster positions and let $x_{i}$ be any real number in the interval $\left[0,L\right]$. We focus on the specific order $x_{1}\leq x_{2}\leq \ldots \leq x_{N_{\mathrm{c}}}$, which is one out of the $N_{\mathrm{c}}!$ possible orderings of coordinates. Therefore, a microstate of the by-now continuum model is going to be represented by a $2N_{\mathrm{c}}$-uple $(n_{1},x_{1}\ldots ,n_{N_{\mathrm{c}}},x_{N_{\mathrm{c}}})$, with $0\leq x_{1}\leq x_{2}\leq \ldots \leq x_{N_{\mathrm{c}}}\leq L$.

Since the clusters are now free to move, the new Hamiltonian is written as:
\begin{equation}
H=\sum_{i=1}^{N_{\mathrm{c}}}\frac{p_{i}^{2}}{2m_{i}}+\frac{1}{2}\epsilon\sum_{i=1}^{N_{\mathrm{c}}}n_{i}(n_{i}-1)+\sum_{i=1}^{N_{\mathrm{c}}-1}u_{a}(x_{i+1}-x_{i})n_{i}n_{i+1}\,,  
\label{4-1}
\end{equation}
where $p_{i}$ and $m_{i}=n_{i}m$ are the momentum and mass of the $i$-th cluster, respectively ($m$ being the mass of an individual particle). In the interaction term, which is again \emph{limited to nearest-neighbors}, the potential $u_{a}$ differs from $u$ for the addition of a suitable hard core of diameter $a$. We need such a modification to ensure that, for the given $u$ potential, a cluster effectively interacts only with its first neighbors; under this assumption, the calculation of the partition function becomes feasible by elementary methods, as we shall see below. Admitting only NN interactions in Eq.\thinspace (\ref{4-1}) actually amounts to cut the $u$ interaction at $r=2a$. For the GEM-$\alpha $ potential with $\alpha\geq 4$, this is clearly allowed for $a>1.1$, but we can take an even smaller $a$ if, in the system configurations with highest Boltzmann's weight under the given thermodynamic conditions, the number of second-neighbor clusters at distance $2a$ is statistically irrelevant (as occurs in GEM-$\alpha$ systems, with $\alpha =4$ or higher, for low temperatures). In other words, replacing $u$ with $u_{a}$ would not alter the statistical properties of the GEM-$\alpha $ model, in so far as the most relevant configurations of model (\ref{4-1}) are those where the distance between NN clusters is around $1.2\div 1.5$.

For reasons which will be apparent below, it is far easier to compute the partition function of model (\ref{4-1}) in an ensemble where the independent variables are $T,P,\mu,N_{\mathrm{c}}$. Observe that $T,P,\mu$ do not exhaust the set of intensive variables for our system since $\mu_{\mathrm{c}}$, the chemical potential conjugate to $N_{\mathrm{c}}$, is also intensive. Denoting $\mathcal{Y}$ the partition function in the $T,P,\mu ,N_{\mathrm{c}}$ representation, the appropriate thermodynamic potential is
\begin{equation}
\mathcal{G}(T,P,\mu ,N_{\mathrm{c}})=-k_{\mathrm{B}}T\ln \mathcal{Y}=\mu _{\mathrm{c}}(T,P,\mu )N_{\mathrm{c}}\,.
\label{4-2}
\end{equation}
However, thermodynamic equilibrium requires that $\mu _{\mathrm{c}}(T,P,\mu)$ must be identically zero\thinspace \cite{Swope,Mladek3,Zhang2,note2} and $\mathcal{G}$ then vanishes exactly. It is worth stressing that the equilibrium condition 
\begin{equation}
\mu _{\mathrm{c}}(T,P,\mu )=0
\label{4-3}
\end{equation}
represents itself an implicit functional relation between $T,P$ and $\mu$, and we will exploit it to express, \textit{e.g.}, $\mu $ as a function of $T$ and $P$.

Statistical mechanics allows to compute $\mathcal{Y}$ from the Hamiltonian. All details of such a calculation are given in the Appendix. The main result is that, similarly to the treatment of Section III, also $\mathcal{Y}(T,P,\mu,N_{\mathrm{c}})$ can be expressed in terms of a transfer matrix ${\mathbb{T}}$ (defined at Eq\,(\ref{a-4})). In the thermodynamic limit, $\mathcal{Y}$ reduces to
\begin{equation}
\mathcal{Y}=\frac{w_{1}^{2}}{L_{0}\Lambda (\beta P)^{2}}\lambda _{\mathrm{max}}^{N_{\mathrm{c}}-1}\,,
\label{4-4}
\end{equation}
where $\lambda _{\mathrm{max}}$ is the maximum eigenvalue of ${\mathbb{T}}$ (and the other, less important, symbols are defined in the Appendix). Substitution of this result into Eq.\,(\ref{4-2}) then yields
\begin{equation}
\mathcal{G}(T,P,\mu ,N_{\mathrm{c}})=-k_{\mathrm{B}}TN_{\mathrm{c}}\ln
\lambda_{\mathrm{max}}+\mathcal{O}(1)\,,
\label{4-5}
\end{equation}
and from Eq.\,(\ref{4-3}) we can conclude that
\begin{equation}
\lambda _{\mathrm{max}}(T,P,\mu )=1\,.
\label{4-6}
\end{equation}
We have thus demonstrated that the equilibrium condition based upon the chemical potential, $\mu_{\mathrm{c}}=0$, is perfectly equivalent to the condition $\lambda_{\mathrm{max}}=1$ for the largest eigenvalue of the transfer matrix ${\mathbb{T}}$.

Before going on, it is worth remarking that the same Eq.\thinspace (\ref{4-6}) can also be obtained by a different route, which is reminiscent of the treatment of 1D multi-component mixtures by Longuet-Higgins\thinspace\cite{LonguetHiggins,Rowlinson,BenNaim}. In this case, let us start from the grand-canonical partition function $\Xi $, defined as 
\begin{equation}
\Xi (T,L,\mu ,\mu _{\mathrm{c}})=\sum_{N_{\mathrm{c}}=1}e^{\beta\mu_{\mathrm{c}}N_{\mathrm{c}}}\sum_{n_{1},\ldots ,n_{N_{\mathrm{c}}}}e^{\beta\mu\sum_{i}n_{i}}\frac{1}{N_{\mathrm{c}}!h^{N_{\mathrm{c}}}}\int_{L^{N_{\mathrm{c}}}}\mathrm{d}^{N_{\mathrm{c}}}p\,\mathrm{d}^{N_{\mathrm{c}}}x\,e^{-\beta H}\,.
\label{4-7}
\end{equation}
Assuming $\mu_{\mathrm{c}}=0$ from the outset, the partition function actually reads: 
\begin{equation}
\Xi (T,L,\mu ,0)=\sum_{N_{\mathrm{c}}=1}^{\infty }\sum_{n_{1},\ldots ,n_{N_{\mathrm{c}}}}e^{\beta \mu \sum_{i}n_{i}}\frac{1}{N_{\mathrm{c}}!h^{N_{\mathrm{c}}}}\int_{L^{N_{\mathrm{c}}}}\mathrm{d}^{N_{\mathrm{c}}}p\,\mathrm{d}^{N_{\mathrm{c}}}x\,e^{-\beta H}\,.
\label{4-8}
\end{equation}
For any number $p$, let us consider the integral: 
\begin{equation}
I=\frac{1}{L_{0}}\int_{0}^{\infty }\mathrm{d}L\,e^{-\beta pL}\Xi (T,L,\mu,0)\,.
\label{4-9}
\end{equation}
Denoting $P(T,\mu)$ the yet unknown system pressure in terms of $T$ and $\mu$, and observing that $\Xi =\exp \{\beta P(T,\mu )L\}$, it follows for any $p>P(T,\mu )$ that: 
\begin{equation}
I=\frac{1}{L_{0}}\int_{0}^{\infty }\mathrm{d}L\,e^{-\beta (p-P(T,\mu ))L}=\frac{1}{\beta L_{0}[p-P(T,\mu )]}\,.
\label{4-10}
\end{equation}
On the other hand, by the same steps leading to Eq.\thinspace (\ref{4-4}) the integral in Eq.\thinspace (\ref{4-10}) can also be more cumbersomely written as: 
\begin{equation}
I=\sum_{N_{\mathrm{c}}=1}^{\infty }\frac{1}{L_{0}\Lambda (\beta p)^{2}}\sum_{n=1}^{\infty }w_{n}^{2}\lambda _{n}^{N_{\mathrm{c}}-1}=\frac{1}{L_{0}\Lambda (\beta p)^{2}}\sum_{n=1}^{\infty }\frac{w_{n}^{2}}{1-\lambda_{n}}\,,
\label{4-11}
\end{equation}
$\{\lambda_{n}\}$ being the set of (real) ${\mathbb{T}}$ eigenvalues listed in decreasing order (notice that all $\lambda _{n}$ are in modulus less than 1 for $p>P(T,\mu)$, since otherwise one of the series in Eq.\,(\ref{4-11}) would not converge, contrary to what implied by Eq.\thinspace (\ref{4-10})). Comparing the last two equations, we see that when $p$ attains the value $P(T,\mu)$ the largest eigenvalue $\lambda _{\mathrm{max}}=\lambda_{1}(T,p,\mu)$ becomes equal to 1, meaning that $P=P(T,\mu)$ is implicitly defined by Eq.\thinspace(\ref{4-6}).

Clearly, the just stated condition would make the thermal properties of the model depend on the exact value of $X=h/\sqrt{2\pi m\epsilon \sigma ^{2}}$, a quantity which sets the value of $\Lambda \sqrt{k_{\mathrm{B}}T/\epsilon }$ relative to $\sigma $ (see Eq.\thinspace (\ref{a-4})). We suspect that this annoying feature of model (4.1) is the price to pay for clusters to be considered as point-like particles. In the following, we take the pragmatic view to choose the $X$ value which produces the best possible agreement with existing MC simulation data for the 1D GEM-4 system.

For $\alpha =4$, we computed $\mu $ and $\rho =(\partial P/\partial \mu)_{T} $ as a function of $P$ in the interval 0.75-0.9, corresponding to the 1-2 transition region, for a number of temperatures between $0.002$ and $0.025$. As a preliminary check, we verified that the $\rho $ values are relatively insensitive to $a$: with either $a=0.5$ or $a=1$ the results were practically the same, suggesting that the distance between NN clusters never falls below 1 in the configurations with the highest statistical weight. This result is very important, since it is then immaterial which potential, either the original $u$ or $u_{a}$ (for $a>0.5$ and NN interactions only), is used for describing the properties of the GEM-4~\cite{note3}.

Our results are plotted in Fig.\thinspace 9 (top panel), for $X=0.01$ and $a=1$. Upon comparing the present results with MC density data for $T=0.02$~\cite{Prestipino3}, we see that the improvement of the continuum model (\ref{4-1}) over the lattice model of Sect.\thinspace III is dramatic (a few $\rho$ curves for the lattice model were reproduced for convenience in the lower panel of Fig.\thinspace 9). In fact, there is a slight dependence of the location of the 1-2 transition on $X$ (all density curves are rigidly shifted to the right as $X$ is reduced), but this effect is smaller for the lower temperatures and actually negligible near $T=0$. However, the crucial point is that, compared to the lattice model of Sect.\thinspace III, the scenario set by the continuum model is qualitatively different: \emph{the 1-2 transition, which for low $T$ is discontinuous in the lattice model, is now turned into a smooth crossover}. This conclusion is consistent with heuristic arguments forbidding any sort of spontaneous symmetry breaking in one dimension~\cite{Peierls,Landau}. One might observe that this outcome does not come as a surprise, and could have been anticipated from the very beginning, since endowing the interaction potential with a hard core restores the validity of van Hove's theorem. Although correct, however, in no way this reasoning diminishes the virtue of a model whose surprising effectiveness, along with the insensitivity of density curves to $a$, make us believe that it is indeed capturing the very essence of the 1D GEM-4 system.

The noteworthy efficacy of the continuum model would cast a shadow on the relevance of the lattice model for the behavior of 1D GEM-4. In other words, the phase transitions discussed in Sect.\thinspace II and III now appear to be an artifice of assuming a strict crystal structure for an originally continuous system of low dimensionality. Nor the simulation results of Ref.~\cite{Prestipino3} can be taken in support of first-order transition behavior. In fact, the density hysteresis loops observed in the simulations appeared to progressively narrow as the MC trajectories became longer and longer, thus rendering those numerical results actually inconclusive.

\section{Conclusions}

Particles interacting through pairwise repulsions bounded at the origin are known to exhibit unconventional thermodynamic and structural behavior, such as reentrant melting, waterlike anomalies, and cluster crystals. Which features are observed in a given case crucially depends on the range of the interaction and the space dimensionality.

In three dimensions, the generalized exponential model of index $\alpha $ (GEM-$\alpha $) shows reentrant melting for $\alpha =2$, whereas it gives rise to a sequence of stable cluster-crystal phases for any $\alpha >2$. In this paper, we have studied the phase behavior of the one-dimensional (1D) GEM-$\alpha $ systems by theoretical means, focusing on the $\alpha =4$ case where numerical-simulation results are available\thinspace \cite{Prestipino3}. MC simulations gave contrasting indications: on the one hand, the low-temperature hysteretic behavior of the number density points to the existence of low-temperature first-order transitions between cluster phases; on the other hand, the systematic narrowing of hysteresis loops, as simulations were made longer and longer, suggests that the apparently singular behavior of the 1D GEM-4 system is actually a finite-size, finite-time artifact.

To clarify the situation, we introduced both a lattice and a continuum cluster model aimed at specifically representing the low-temperature behavior of the original GEM system. While the lattice system exhibited a sequence of low-temperature first-order transitions between cluster phases of increasing occupancy, as evidenced by the discontinuous jumps of the number density as a function of pressure, the more accurate model defined on the continuum only showed a series of smooth crossovers between cluster ``phases'' devoid of separate individuality. When comparing the results for the continuum cluster model with MC simulation data for $\alpha=4$ we found a rather good agreement, thus indicating that strict phase boundaries are likely absent in the 1D GEM-4 system, and even more so in the 1D GEM-$\alpha$ with $\alpha>4$.

\acknowledgments We are grateful to Andr\'{e}s Santos (University of Extremadura, Spain) for useful discussions. N. T. wishes to thank Universit\`{a} Ca' Foscari Venezia for his post-doc position.

\appendix
\section{Calculation of $\mathcal{Y}(T,P,\protect\mu ,N_{\mathrm{c}})$}
\setcounter{equation}{0}
\renewcommand{\theequation}{A.\arabic{equation}}

The kinetic part of $\mathcal{Y}$ is immediately calculated after noting that 
\begin{equation}
\frac{1}{h}\int_{-\infty }^{\infty }\mathrm{d}p_{i}\exp \left\{ -\frac{\beta p_{i}^{2}}{2m_{i}}\right\} =\frac{\sqrt{n_{i}}}{\Lambda }\,,
\label{a-1}
\end{equation}
where $\Lambda=h/\sqrt{2\pi mk_{\mathrm{B}}T}$ is the thermal wavelength of an individual particle. The partition function then becomes:
\begin{eqnarray}
\mathcal{Y}(T,P,\mu ,N_{\mathrm{c}}) &=&\sum_{n_{1},\ldots ,n_{N_{\mathrm{c}}}}e^{\beta \mu \sum_{i}n_{i}}\frac{1}{L_{0}}\int_{0}^{\infty }\mathrm{d}L\,e^{-\beta PL}\frac{1}{N_{\mathrm{c}}!h^{N_{\mathrm{c}}}}\int_{L^{N_{\mathrm{c}}}}\mathrm{d}^{N_{\mathrm{c}}}p\,\mathrm{d}^{N_{\mathrm{c}}}x\,e^{-\beta H}
\notag \\
&=&\sum_{n_{1},\ldots ,n_{N_{\mathrm{c}}}}e^{\beta \mu\sum_{i}n_{i}}e^{-(\beta \epsilon
/2)\sum_{i}n_{i}(n_{i}-1)}e^{(1/2)\sum_{i}\ln n_{i}}
\notag \\
&\times &\frac{1}{L_{0}}\int_{0}^{\infty }\mathrm{d}L\,e^{-\beta PL}\frac{1}{N_{\mathrm{c}}!\Lambda^{N_{\mathrm{c}}}}\int_{L^{N_{\mathrm{c}}}}\mathrm{d}x_{1}\cdots \mathrm{d}x_{N_{\mathrm{c}}}\,e^{-\beta\sum_{i}u_{a}(x_{i+1}-x_{i})n_{i}n_{i+1}}\,,
\label{a-2}
\end{eqnarray}
where $L_{0}$ is an arbitrary length.

In order to compute the integrals in Eq.\thinspace (\ref{a-2}), we shall proceed \emph{\`{a} la} Takahashi~\cite{Takahashi}, writing the inner integral as a chain of convolutions whose Laplace transform is easily computed (that is why working at fixed $P$ is more convenient). The result
is (cf. appendix A of Ref.\thinspace \cite{Sadr-Lahijany}):
\begin{eqnarray}
&&\frac{1}{L_{0}}\int_{0}^{\infty }\mathrm{d}L\,e^{-\beta PL}\frac{1}{N_{\mathrm{c}}!\Lambda ^{N_{\mathrm{c}}}}\int_{L^{N_{\mathrm{c}}}}\mathrm{d}x_{1}\cdots \mathrm{d}x_{N_{\mathrm{c}}}\,e^{-\beta\sum_{i}u_{a}(x_{i+1}-x_{i})n_{i}n_{i+1}}
\notag \\
&=&\frac{1}{L_{0}\Lambda ^{N_{\mathrm{c}}}}\int_{0}^{\infty }\mathrm{d}L\,e^{-\beta PL}\int_{x_{1}<\ldots <x_{N_{\mathrm{c}}}}\mathrm{d}x_{1}\cdots\mathrm{d}x_{N_{\mathrm{c}}}\,e^{-\beta\sum_{i}u_{a}(x_{i+1}-x_{i})n_{i}n_{i+1}}
\notag \\
&=&\frac{1}{L_{0}\Lambda (\beta P)^{2}}\prod_{i=1}^{N_{\mathrm{c}}-1}\int_{0}^{\infty }\frac{\mathrm{d}s}{\Lambda }\,e^{-\beta Ps}e^{-\beta u_{a}(s)n_{i}n_{i+1}}\,.
\label{a-3}
\end{eqnarray}
Notice that Eq.\thinspace(\ref{a-3}) is only valid under the assumption of \emph{open boundary conditions}. Upon defining the transfer matrix ${\mathbb{T}}$ as the following symmetric matrix,
\begin{equation}
{\mathbb{T}}_{n^{\prime }}^{n}=e^{\beta \mu (n+n^{\prime })/2}e^{-\beta\epsilon (n^{2}+n^{\prime 2}-n-n^{\prime })/4}e^{(1/4)(\ln n+\ln n^{\prime})}\int_{a}^{\infty }\frac{\mathrm{d}s}{\Lambda }\,e^{-\beta Ps}e^{-\beta u(s)nn^{\prime }}\,,
\label{a-4}
\end{equation}
Eq.\thinspace(\ref{a-2}) is eventually rewritten as: 
\begin{eqnarray}
\mathcal{Y} &=&\frac{1}{L_{0}\Lambda (\beta P)^{2}}\sum_{n_{1},\ldots ,n_{N_{\mathrm{c}}}}e^{\beta \mu n_{1}/2}e^{-\beta \epsilon n_{1}(n_{1}-1)/4}e^{(1/4)\ln n_{1}}{\mathbb{T}}_{n_{2}}^{n_{1}}{\mathbb{T}}_{n_{3}}^{n_{2}}\cdots {\mathbb{T}}_{n_{N_{\mathrm{c}}}}^{n_{N_{\mathrm{c}}-1}}
\notag \\
&\times &e^{\beta \mu n_{N_{\mathrm{c}}}/2}e^{-\beta \epsilon n_{N_{\mathrm{c}}}(n_{N_{\mathrm{c}}}-1)/4}e^{(1/4)\ln n_{N_{\mathrm{c}}}}=\frac{1}{L_{0}\Lambda (\beta P)^{2}}\mathbf{v}^{\mathrm{T}}{\mathbb{T}}^{N_{\mathrm{c}}-1}\mathbf{v}\,,
\label{a-5}
\end{eqnarray}
where \textbf{v} is the $N_{\mathrm{c}}$-dimensional column vector with components 
\begin{equation}
v_{i}=e^{\beta \mu n_{i}/2}e^{-\beta \epsilon n_{i}(n_{i}-1)/4}e^{(1/4)\ln n_{i}}\,.
\label{a-6}
\end{equation}
Equation (\ref{a-5}) is an exact result, which involves $N_{\mathrm{c}}$ in a complicate manner. However, in the $N_{\mathrm{c}}\rightarrow\infty$ limit Eq.\thinspace (\ref{a-5}) is radically simplified, as we are going to show soon (observe that, when $N_{\mathrm{c}}\rightarrow \infty $ also $N=\sum_{i=1}^{N_{\mathrm{c}}}n_{i}\rightarrow \infty $).

Let ${\mathbb{R}}$ be the orthogonal matrix having the ${\mathbb{T}}$ eigenvectors for columns. Then
\begin{equation}
\mathbf{v}^{\mathrm{T}}{\mathbb{T}}^{N_{\mathrm{c}}-1}\mathbf{v}=\mathbf{w}^{\mathrm{T}}{\mathbb{T}}^{\prime N_{\mathrm{c}}-1}\mathbf{w}\,\,\,\,\,\,\mathrm{with}\,\,\,\,\,\,\mathbf{w}={\mathbb{R}}^{-1}\mathbf{v}\,\,\,\mathrm{and}\,\,\,{\mathbb{T}}^{\prime }={\mathbb{R}}^{-1}{\mathbb{T}}{\mathbb{R}}\,.
\label{a-7}
\end{equation}
Called $\lambda_{\mathrm{max}}$ the maximum eigenvalue of ${\mathbb{T}}$, in the thermodynamic limit the expression of $\mathcal{Y}$ simplifies to
\begin{equation}
\mathcal{Y}=\frac{w_{1}^{2}}{L_{0}\Lambda (\beta P)^{2}}\lambda _{\mathrm{max}}^{N_{\mathrm{c}}-1}\,,
\label{a-8}
\end{equation}
which is the same as Eq.\,(\ref{4-4}).

\newpage 
\begin{figure}[tbp]
\centering
\includegraphics[width=14cm]{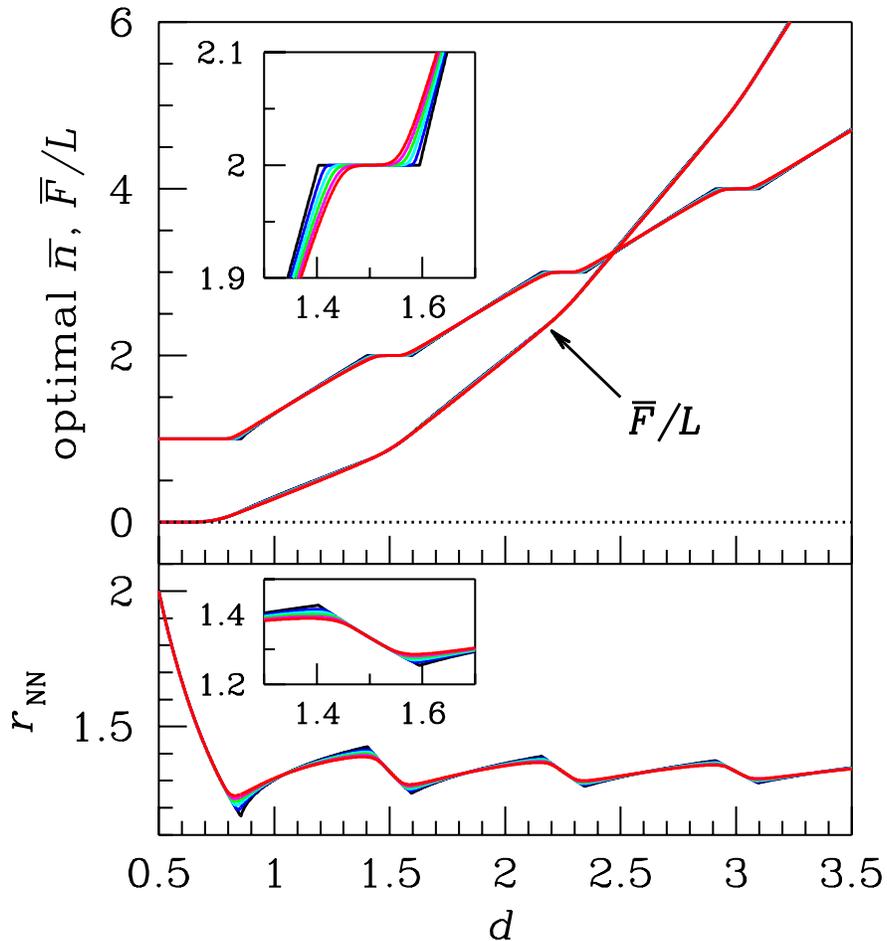}
\caption{(Color online). 1D GEM-4, mean-field results for six temperatures, $T=0,0.01,\ldots,0.05$, marked as black, blue, cyan, green, magenta, and red lines, respectively. We here show results deriving from the optimization of $\overline{n}$ only (top: best $\overline{n}$ and $\overline{F}/L$ values; bottom: lattice spacing). Each inset shows a magnification of the $d$ interval from 1.3 to 1.7 (top: $\overline{n}$; bottom: lattice spacing).}
\label{fig1}
\end{figure}

\begin{figure}[tbp]
\centering
\includegraphics[width=14cm]{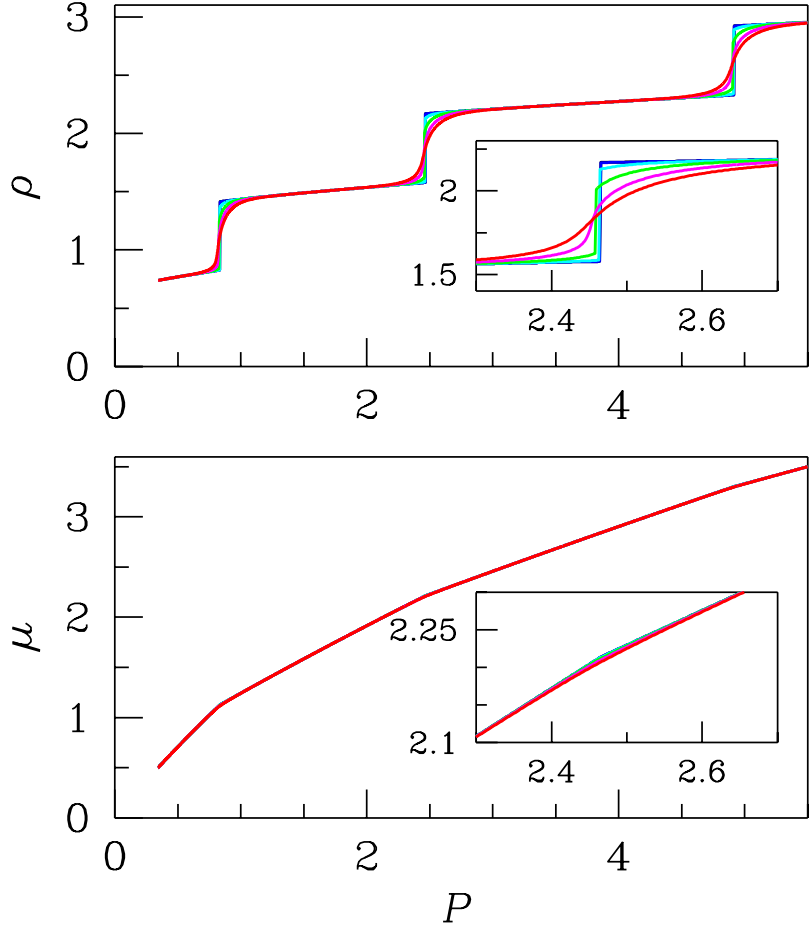}
\caption{(Color online). 1D GEM-4, mean-field results for the same temperatures considered in Fig.\,1. We here plot the density $\protect\rho$ (top) and the chemical potential $\protect\mu$ (bottom) as a function of the pressure $P$, as resulting from the optimization of both $\overline{n}$ and $d$ (only the first three transitions are shown). In each inset, we show a magnification of the $P$ interval from 2.3 to 2.7, which makes us better appreciate the effect of heating the system: while the transition from the 2-cluster to the 3-cluster phase is still first-order for $T=0.03$, it has by now become a sharp (but smooth) crossover for $T=0.04$.}
\label{fig2}
\end{figure}

\begin{figure}[tbp]
\centering
\includegraphics[width=14cm]{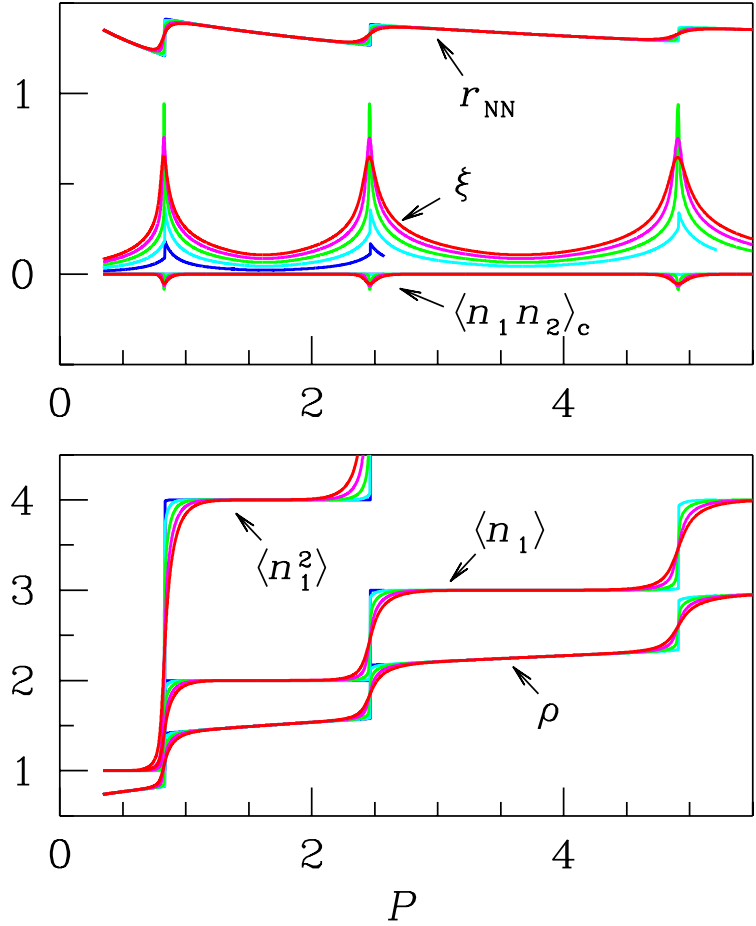}
\caption{1D GEM-4, transfer-matrix results for various temperatures ($T=0,0.01,\ldots,0.05$, same colors as in Figs.\,1 and 2). In the top panel, the average NN distance, the correlation length, and the connected NN correlation are plotted as a function of pressure. In the bottom panel, the average cluster size $\langle n_1\rangle$, the average squared occupancy, and the density are shown.}
\label{fig3}
\end{figure}

\begin{figure}[tbp]
\centering
\includegraphics[width=14cm]{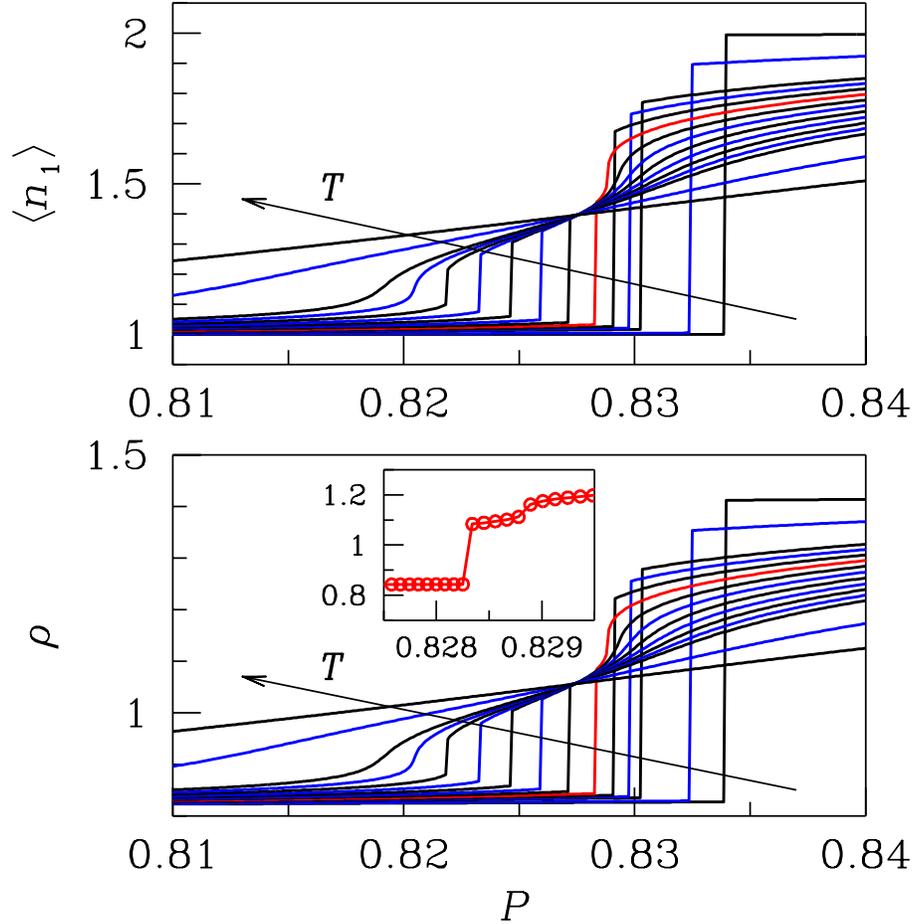}
\caption{1D GEM-4, transfer-matrix results for various temperatures ($T=0.01,0.02,0.025,0.026,\ldots,0.034,0.035,0.04,0.05$; for the sake of clarity, the line colors are alternatively black and blue, except for $T=0.028$ -- red curves -- where the density shows two distinct jumps at two nearby pressures): average cluster size (top panel) and number density (bottom panel) near the transition from the 1-cluster to the 2-cluster phase. In the inset, we show a magnification of the density for $T=0.028$, which clearly signals the existence of two distinct steps (the raw data are the dots, joined by straight-line segments to help the eye).}
\label{fig4}
\end{figure}

\begin{figure}[tbp]
\centering
\includegraphics[width=14cm]{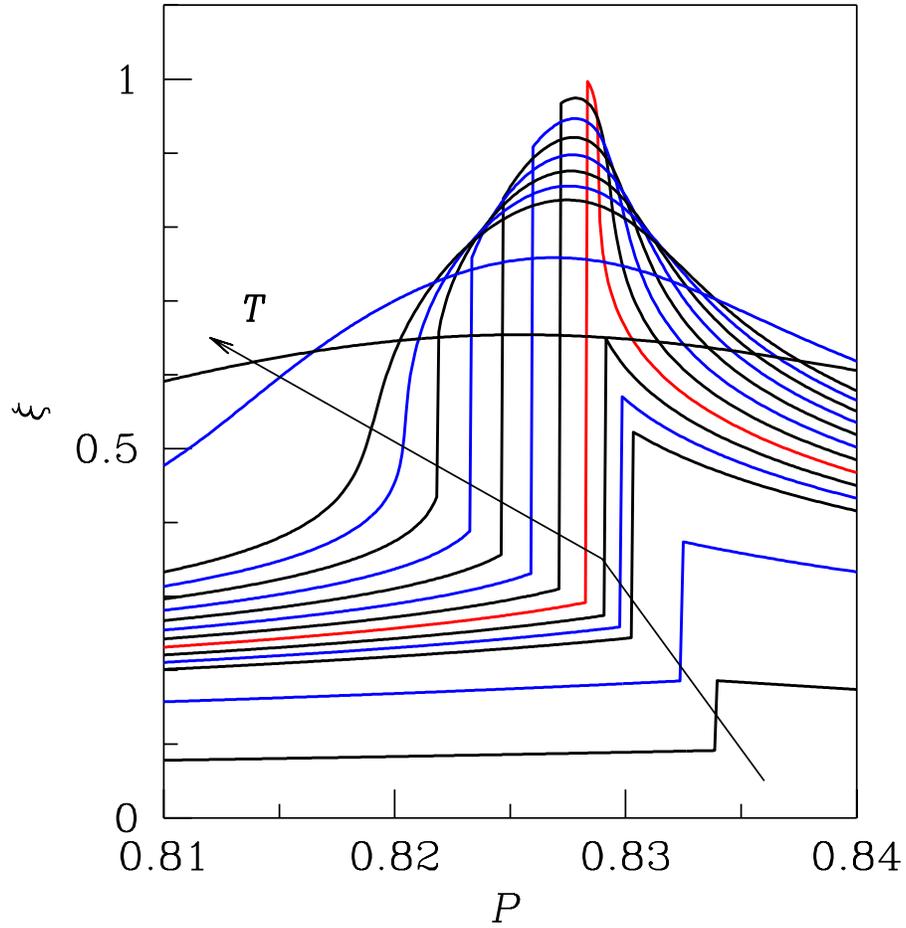}
\caption{1D GEM-4, transfer-matrix results for various temperatures (the same as in Fig.\,4): correlation length $\protect\xi$ near the transition from the 1-cluster to the 2-cluster phase.}
\label{fig5}
\end{figure}

\begin{figure}[tbp]
\centering
\includegraphics[width=14cm]{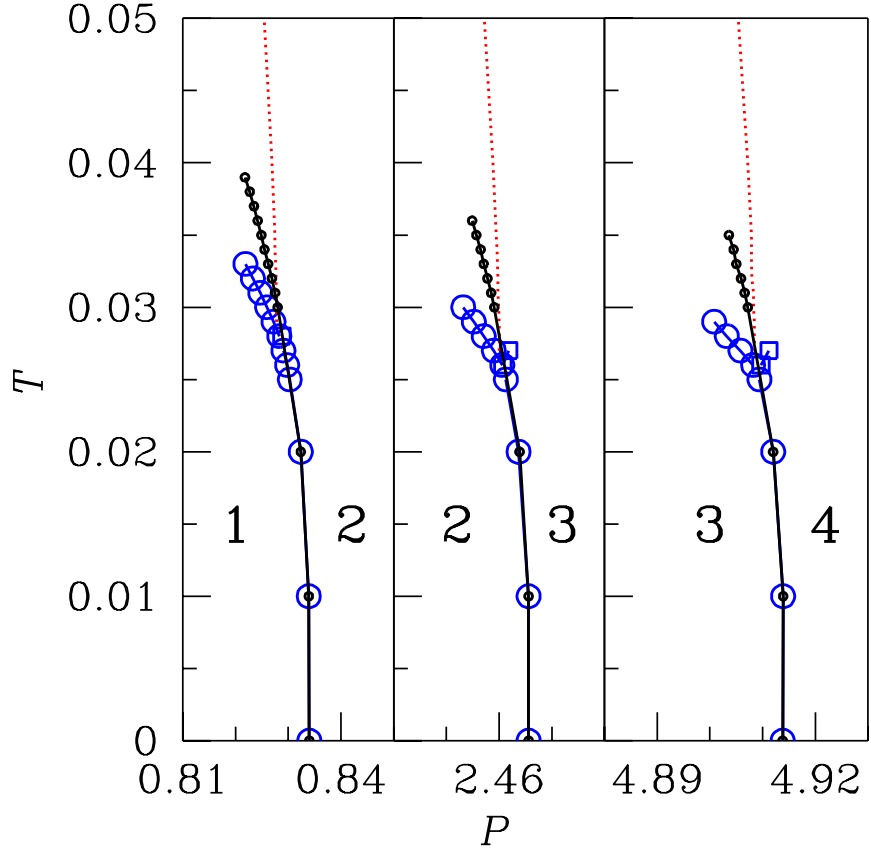}
\caption{1D GEM-4 phase diagram on the $P$-$T$ plane: Exact transfer-matrix results (open blue circles and squares) vs. mean-field results (black dots). The exact Widom line is also shown (dotted red line).}
\label{fig6}
\end{figure}

\begin{figure}[tbp]
\centering
\includegraphics[width=14cm]{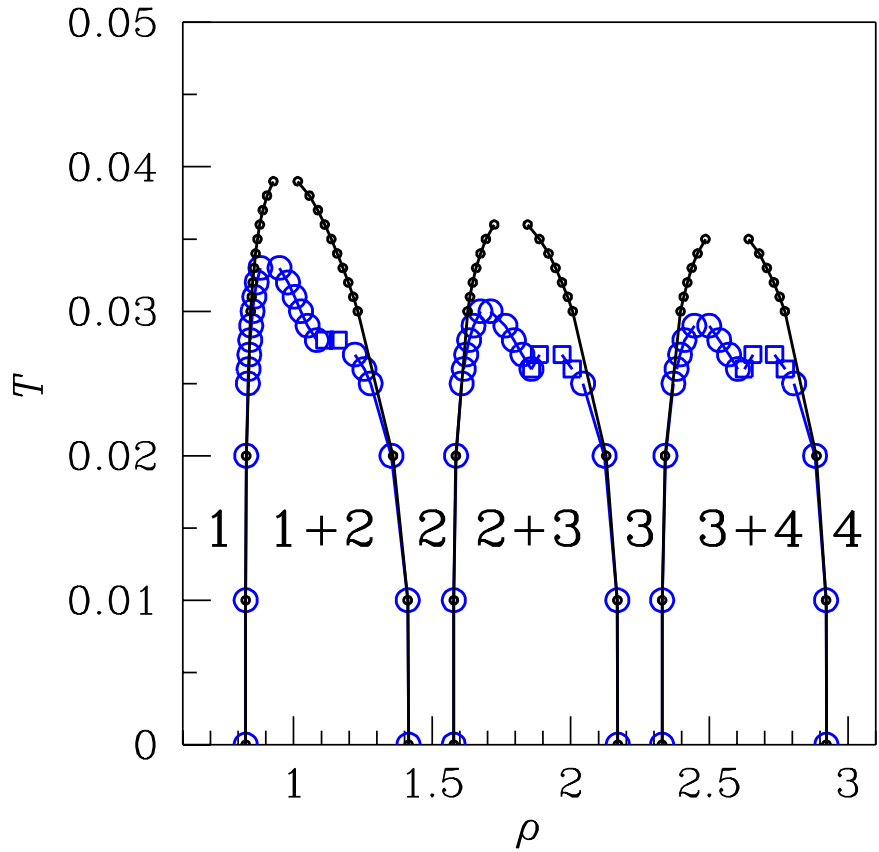}
\caption{1D GEM-4 phase diagram on the $\protect\rho$-$T$ plane: Exact transfer-matrix results (open blue circles and squares) vs. mean-field results (black dots). }
\label{fig7}
\end{figure}

\begin{figure}[tbp]
\centering
\includegraphics[width=14cm]{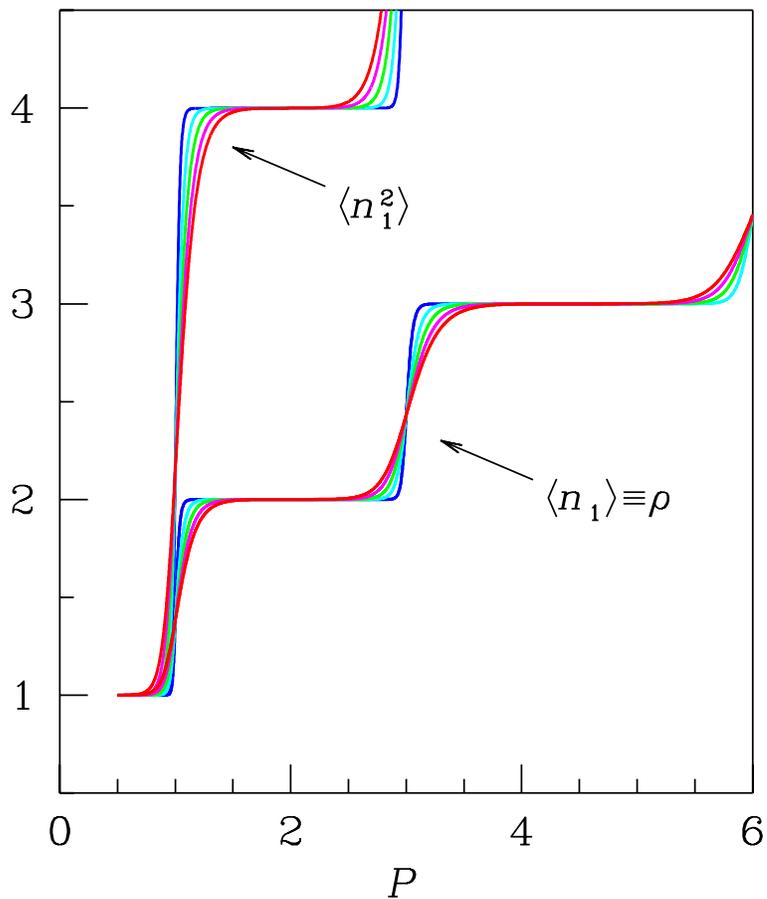}
\caption{1D GEM-$\infty$ (PSM), transfer-matrix results for various temperatures ($T=0,0.01,\ldots,0.05$, same colors as in Figs.\,1, 2, and 3): average cluster size $\langle n_1\rangle$ (which coincides with the density) and average squared size $\langle n_1^2\rangle$. In the $\protect\alpha\rightarrow\infty$ limit, the discontinuous jumps of the density at the lowest temperatures have disappeared, leaving a sharp but continuous crossover at all non-zero temperatures.}
\label{fig8}
\end{figure}

\begin{figure}[tbp]
\centering
\includegraphics[width=12cm]{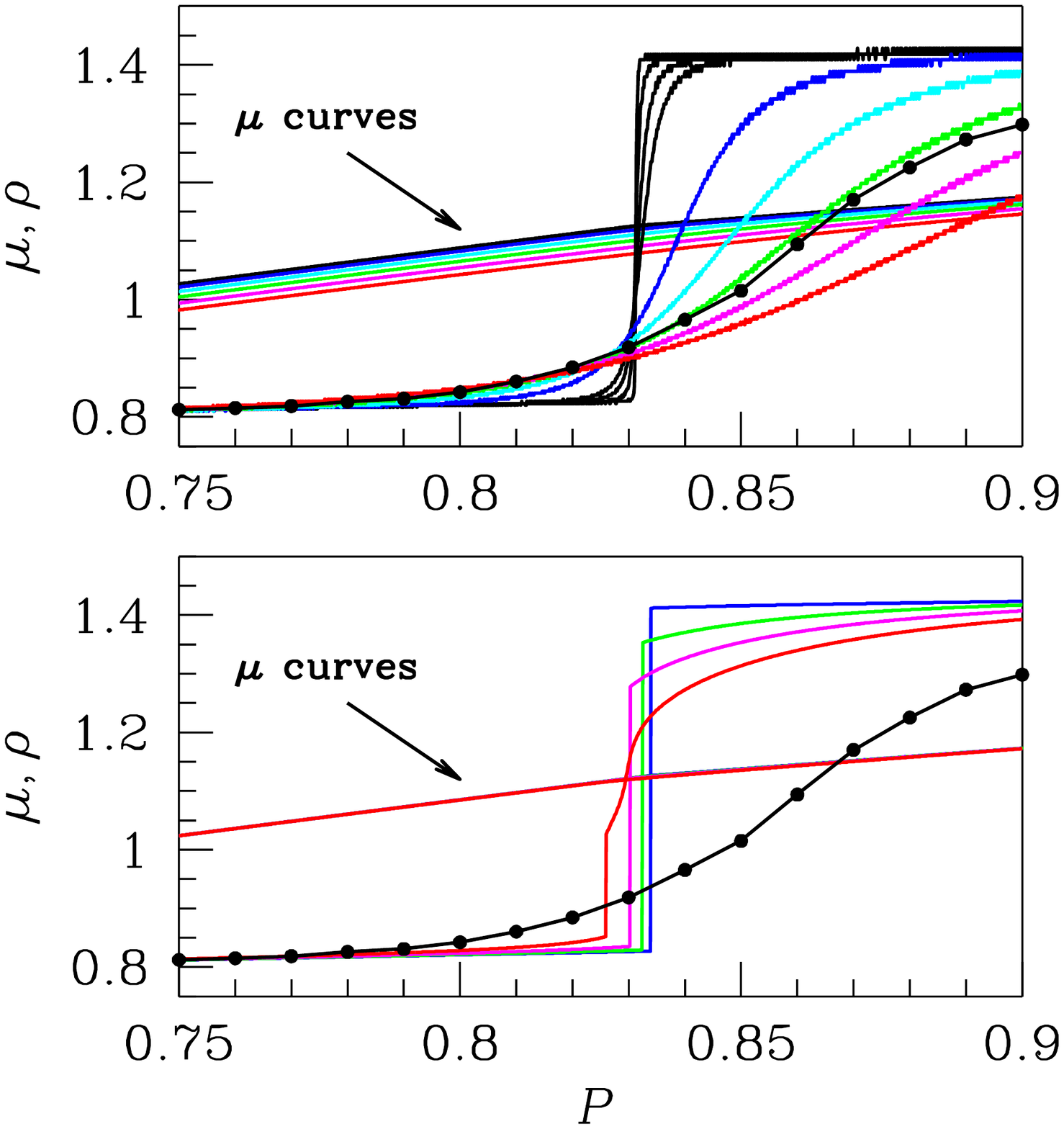}
\caption{1D GEM-4, comparison between $\protect\mu$ and $\protect\rho$ results for the continuous model (Section IV) and for the lattice model (Section III) in the 1-2 transition region (for $X\equiv h/\protect\sqrt{2\protect\pi m\protect\epsilon\protect\sigma^2}=0.01$ and $a=1$). Top: continuous model ($T=0.002,0.003,0.004,0.005$, black; $T=0.01$, blue; $T=0.015$, cyan; $T=0.02$, green; $T=0.025$, magenta; $T=0.03$, red). Observe that even for such a low temperature as $T=0.002$ the density varies smoothly across the ``transition''. In the same panel, the black dots are $T=0.02$ density data from Ref.\,\protect\cite{Prestipino3}. Bottom: lattice model ($T=0.01$, blue; $T=0.02$, green; $T=0.025$, magenta; $T=0.03$, red).}
\label{fig9}
\end{figure}

\begin{thebibliography}{99}
\bibitem{note3} If $a$ becomes too small the profile of the density curves will change appreciably, meaning that the entropic contribution from configurations with many close-by clusters becomes overwhelming (notice that the energy will not increase likewise since clusters at a distance higher than $2a$ do not interact); however, this situation does not seem to correspond to the GEM-4 behavior at low temperature~\cite{Prestipino3}.
\end{thebibliography}
\end{document}